\documentclass[a4paper, 11pt]{article}

\usepackage[affil-it]{authblk} 
\usepackage{etoolbox}
\usepackage{lmodern}

\makeatletter
\patchcmd{\@maketitle}{\LARGE \@title}{\fontsize{16}{16}\selectfont\@title}{}{}
\makeatother

\renewenvironment{abstract}
 {\small
  \begin{center}
  \bfseries \abstractname\vspace{-.5em}\vspace{0pt}
  \end{center}
  \list{}{%
    \setlength{\leftmargin}{5mm}% <---------- CHANGE HERE
    \setlength{\rightmargin}{\leftmargin}%
  }%
  \item\relax}
 {\endlist}

\usepackage{lineno}
\usepackage{amssymb}
\usepackage[nodots]{numcompress}
\usepackage{amssymb,latexsym,amsmath, graphicx, amsfonts, caption, color, verbatim}

\usepackage{subcaption}

\title{Surrogacy of progression free survival for overall survival in metastatic breast cancer studies: meta-analyses of published studies}
\author{\vspace{0.2in} Madan G. Kundu and Suddhasatta Acharyya  \\
Novartis Pharmaceutical Corporation\\
East Hanover, NJ, USA }

\date{}
\begin{document}

\maketitle

%------------------------------------------------------------------------------------------------------------------------------
%                 ABSTRACT
%------------------------------------------------------------------------------------------------------------------------------
\begin{abstract}
\textbf{Purpose:} PFS is often used as a surrogate endpoint for OS in metastatic breast cancer studies. We have evaluated the association of treatment effect on PFS with significant HR$_{OS}$ (and how this association is affected by other factors) in published prospective metastatic breast cancer studies.

\textbf{Methods:} A systematic literature search in PubMed identified prospective metastatic breast cancer studies. Treatment effects on PFS were determined  using hazard ratio (HR$_{PFS}$), increase in median PFS ($\Delta$MED$_{PFS}$) and \% increase in median PFS (\%$\Delta$MED$_{PFS}$). Diagnostic accuracy of PFS measures (HR$_{PFS}$, $\Delta$MED$_{PFS}$ and \%$\Delta$MED$_{PFS}$) in predicting significant HR$_{OS}$ was assessed using receiver operating characteristic (ROC) curves and classification tree approach (CART). 

\textbf{Results:} Seventy-four cases (i.e., treatment to control comparisons) from 65 individual publications were identified for the analyses. Of these, 16 cases reported significant treatment effect on HR$_{OS}$ at 5\% level of significance. Median number of deaths reported in these cases were 153. Area under the ROC curve (AUC) for diagnostic measures as HR$_{PFS}$, $\Delta$MED$_{PFS}$ and \%$\Delta$MED$_{PFS}$ were 0.69, 0.70 and 0.75, respectively. Classification tree results identified \%$\Delta$MED$_{PFS}$ and number of deaths as diagnostic measure for significant HR$_{OS}$. Only 7.9\% (3/39) cases with $\Delta$MED$_{PFS}$ shorter than 48.27\% reported significant HR$_{OS}$. There were 7 cases with $\Delta$MED$_{PFS}$ of 48.27\% or more and number of deaths reported as 227 or more -- of these 5 cases reported significant HR$_{OS}$.

\textbf{Conclusion:} \%$\Delta$MED$_{PFS}$ was found  to be a better diagnostic measure for predicting significant HR$_{OS}$. Our analysis results also suggest that consideration of total number of deaths may further improve its diagnostic performance. Based on our study results,  the studies with 50\% improvement in median PFS are more likely to produce significant HR$_{OS}$ if the total number of OS events at the time of analysis is 227 or more.

\end{abstract}


%------------------------------------------------------------------------------------------------------------------------------
%                 INTRODUCTION
%------------------------------------------------------------------------------------------------------------------------------
\section*{Introduction}\label{sec:intro}
As per national cancer institute, in the U.S., breast cancer is the second most common non-skin cancer and the second leading cause of cancer-related deaths in women; and, therefore, there has always been a high demand for novel breast cancer therapies.  At the time of preparing this manuscript\textcolor{black}{, based on ClinicalTrial.gov search \cite{clintrialgov},} 175 phase III breast cancer studies were actively recruiting patients. For breast cancer therapies, the main goal is to improve overall survival (OS) and quality of life \cite{saad2010progression, smith2006goals}. US FDA guideline \cite{FDA2007cancer}  states that ``[overall] survival is considered the most reliable cancer endpoint''. Due to the advancement in metastatic breast cancer management and therapies, there has been marked improvement in OS in breast cancer patients in the last few decades. \textcolor{black}{Conequently, patients need to be followed-up for longer period of time to observe sufficient  number of OS events (i.e., deaths) \cite{booth2012progression} before treatment effect on OS can be evaluated statistically. Further, as many patients switch to second line (and beyond) therapies upon progression, the OS time may be influenced by post-progression therapy.} For these reasons, surrogate endpoints such as progression-free survival (PFS) or objective response rate (ORR) are being increasingly used for accelerated approvals, with PFS being the one used most often \cite{saad2010progression}. The basis for using PFS as surrogate endpoint for OS is as follows: cancer progression represents an ominous march toward death from malignancy. Hence, the longer it takes for the cancer to progress, the longer a patient will live \cite{venook2014progression}.  In general, PFS has not been statistically validated for surrogacy of OS yet \textcolor{black}{in breast cancer studies}\cite{saad2010progression, FDA2007cancer}. \textcolor{black}{Reported results  regarding association between Hazard ratio of PFS and OS in the metastatic breast cancer studies are mixed: For example, Hackshaw {\it et al.} \cite{hackshaw2005surrogate} found a correlation of 0.87; Burzykowski {\it et al.} \cite{burzykowski2008evaluation} reported correlation of 0.48; Michiels {\it et al.} \cite{michiels2016progression} reported $R^2$ (i.e. proportion of the variance in the true endpoint that is explained by the surrogate endpoint) as $0.51$.}\\ 

According to Prentice's definition\cite{prentice1989surrogate}, in order for PFS to be a ``statistically validated'' surrogate endpoint for OS, ``test for null hypothesis of no treatment effect in PFS" should be a valid ``test for null hypothesis of no treatment effect in OS". The test for treatment effect on OS is carried out by testing HR$_{OS}$=1, where HR$_{OS}$ is the hazard ratio (HR) of OS. However, many randomized clinical trials failed to demonstrate significant treatment effect in OS despite demonstrating significant treatment effect in PFS. The current project attempts to investigate the trial level surrogacy in breast cancer studies from a diagnostic testing perspective using nonparametric  approaches.  It is important to note that our investigation differs from previous investigations \cite{hackshaw2005surrogate, burzykowski2008evaluation, bruzzi2005objective,  miksad2008progression, sherrill2008relationship, ng2008correlation, sherrill2012review, amiri2016association} based on meta-analytic methods, where the primary purpose was to examine the strength of treatment effect on PFS to predict treatment effect on OS at trial level. The definition of trial level surrogacy in the current investigation is intuitive and aligned with the ultimate question that all stakeholders, regulators in particular, are often seeking an answer to, from a phase III cancer clinical trial -- Is there a statistically significant OS benefit in the new treatment that is discernible from the data on progression-free survival (PFS) in metastatic breast cancer studies? \textcolor{black}{This definition of trial level surrogacy was also cosidered by Burzykowski and Buyse \cite{burzykowski2006surrogate} as it can be useful to estimate the  ``Surrogate threshold effect''. Surrogate threshold effect can be defined as the minimum treatment effect on PFS measure that is required to predict statistically significant HR$_{OS}$}.\\

\textcolor{black}{Our goal was to evaluate the trial level surrogacy of PFS for OS solely based on published clinical trial results. 
%Previously, trial level surrogacy of PFS for OS based on published studies using parametric modeling \cite{burzykowski2008evaluation, buyse2000validation} have been proposed.
Burzykowski {\it et al.} \cite{burzykowski2008evaluation} evaluated trial level surrogacy by fitting  simple (log-) linear regression analysis to model HR$_{OS}$ with ratio of median PFS time and then used $R^2$  as a measure of trial level surrogacy.  Buyse {\it et al.} \cite{buyse2000validation} proposed to estimate trial level surrogacy using $R^2$ as well, but in a more sophisticated way using trial specific random effects. These methods make various model assumptions such as PFS and OS are linearly associated \cite{burzykowski2008evaluation} or some distributional assumption \cite{buyse2000validation}). As Venook and Tabernero \cite{venook2014progression} have pointed out association of PFS with OS may be complicated in today's era and, therefore, a simplified linear model  may not be sufficient to describe the association. Further, the use of $R^2$ is heavily impacted by the presence of outlier \cite{buyse2007progression}. Another problem related to $R^2$ is the difficulty in interpreting its value \cite{burzykowski2006surrogate}. For these reasons, we  have adopted non-parametric approaches to evaluate the trial level surrogacy which, unlike parametric methods, do not require to make distributional assumptions or to pre-specify the from of the association. The advnatges of non-parametric methods are that these methods are  completley data-driven and free from model assumptions. Consequently, non-parametric methods have obvious advantage of producing results which are solely based on observed data and are not dependent on unverifiable model assumptions.  Non-parametric methods can be also useful (a) to find out which PFS measure is relatively more  important in predicting significant HR$_{OS}$, (b) to study the influence of other factors (e.g., sample size and total number of events) on the association of PFS measure and significant HR$_{OS}$ as, for example, the power for statistical test of HR$_{OS}$ is a function of total number of OS events, and (c) to estimate surrogate threshold effect.} Results from non-parametric  methods  are often easy to interpret, and allow  granular visualization of the results. For this project, breast cancer studies were our focus, but the similar investigation can be carried out for other indications as well. Throughout the article, (unless otherwise mentioned), `statistically significant' would imply that the significance was in favor of the treatment.

\section*{Methods}

\subsection*{Literature search}

A systematic literature search in PubMed (July 2015) was performed to identify published prospective studies on metastatic breast cancer research with both PFS and OS comparison results reported. \textcolor{black}{The search syntax used was as follows: ``(((Breast Cancer[Title]) AND Randomized[Title/Abstract]) AND Progression free survival[Text Word]) AND Overall survival[Text Word]''.} 
%(i) title includes the phrase ``breast cancer'',  (ii) the term ``randomized'' is in title or abstract, (iii) the phrase ``progression free survival'' and ``overall survival'' are mentioned in the text [criteria: Breast cancer (in Title), Randomized (in Title or Abstract), Progression free survival (in Text Word)  and Overall survival (in Text Word)]. 
The PubMed search returned 181 publications between Jul-2000 and Jul-2015. Many of these studies were systematic literature review or meta-analyses and hence dropped. Further, studies with either PFS or OS not reported were also excluded. We were able to find 64 individual prospective studies \cite{ref1, ref2, ref4, ref5, ref6, ref7, ref8, ref9, ref10, ref11, ref12, ref13, ref14, ref15, ref17, ref18, ref19, ref20, ref21, ref22, ref23, ref24, ref25, ref26, ref27, ref28, ref29, ref30, ref31, ref32, ref33, ref34, ref35, ref35a, ref36, ref37, ref38, ref39, ref40, ref41, ref42, ref43, ref44, ref45, ref46, ref47, ref48, ref49, ref50, ref51, ref52, ref53, ref54, ref54a,  ref55, ref56, ref57, ref58, ref59, ref60, ref61, ref62, ref63, ref64} where both PFS and OS comparison results were reported. \textcolor{black}{In addition, in one publication \cite{ref54b}, instead of PFS, time to progression (TTP) was reported and that study was included. Therefore, we had total of 65 publications for the meta-analyses.}  

\begin{table}
\begin{center}
\caption{Summary of publications by journal and year } \label{Table1}
\begin{tabular}
{l|ccccc|r}
\hline
Journal of Publication &2014-2015&2012-2013&2010-2011&2005-2009&2000-2004&Total\\
\hline
Journal of Clinical Oncology&7&9&4&5&3&28\\
Breast Cancer Research &&&&&&\\
Treatment&3&2&4&2&&11\\
Annals of Oncology&5&3&&&1&9\\
Cancer&1&&1&1&&3\\
Clinical Breast Cancer&&&3&&&3\\
Clinical Cancer Research&&2&&&&2\\
Others&1&4&1&2&1&9\\
\hline
Total& 17 & 20 & 13 & 10 & 5 & 65\\
\hline
\end{tabular}
\end{center}
\end{table}

\subsection*{Data extraction}
Of the 65 selected publications, in seven prospective studies \cite{ref7, ref8, ref9, ref39, ref41, ref43, ref64}, two pairs of treatment-to-control comparisons were reported and in one prospective study \cite{ref18}, three pairs of treatment-to-control comparisons were reported. Therefore, we had total of 74 treatment-to-control comparison available for the meta-analyses. For each treatment-to-control comparison, the following information were extracted: randomization status, blinding status (open or blinded),  total sample size (treatment plus control), total number of events (treatment plus control), median PFS, median OS, HR (hazard ratio) in PFS (HR$_{PFS}$), HR in OS HR$_{OS}$, reported p-value (or significance status) for HR$_{PFS}$ and reported p-value (or significance status) for HR$_{OS}$. In case both local and central PFS assessments were reported, the one which was reported as primary endpoint was considered. %In most cases Treatment to Control HR were considered for the meta-analysis, except for ?? studies where control arm had better OS compared to investigational drug and for these studies control to treatment HR ratio were used.

\subsection*{Statistical methods}

Treatment effect on PFS was determined using the following measures: hazard ratio (HR$_{PFS}$), increase in median PFS ($\Delta$MED$_{PFS}$) and \% increase in median PFS (\%$\Delta$MED$_{PFS}$). All three measures were used as diagnostic tools for predicting statistically significant HR$_{OS}$ in favor of treatment (yes/no).\\

\textcolor{black}{We have assessed the trial level surrogacy of PFS for OS by evaluating the diagnostic accuracy of these comparative PFS measures to predict statistically significant HR$_{OS}$. Diagnostic accuracy of comparative PFS measures (HR$_{PFS}$, $\Delta$MED$_{PFS}$ and \%$\Delta$MED$_{PFS}$) in predicting significant HR$_{OS}$ was assessed using receiver-operating characteristic (ROC) curve \cite{zhou2009statistical}, and classification tree (using CART algorithm \cite{breiman1984classification}) approach.} Empirical ROC curves were drawn plotting the true positive rate (proportion of correct prediction of significant HR$_{OS}$ based on comparative PFS measure among those reporting significant treatment effect on HR$_{OS}$) against the false positive rate (proportion of wrong prediction of significant HR$_{OS}$ based on PFS measure among those reported non- significant treatment effect on HR$_{OS}$). \textcolor{black}{ True positive rate, and false positive rate were obtained at each unique value of comparative PFS measures. For a given unique value of $x$, if comparative PFS measure was greater than or equal to $x$, then it was predicted that HR$_{OS}$ will be signifacant; otherwise not.} The accuracy of the diagnostic measure was assessed by numerically computing the area under ROC curve (AUC), with larger AUC implying better accuracy. Optimal cut-off points based on ROC curve were identified according to Youden's index\cite{youden1950index}. According to Youden's criteria a optimum cut-off point for prediction of significant HR$_{OS}$ would be one that maximizes the difference between true positive rate and false positive rate.\\

We have utilized classification tree to answer following questions: (a) which trial level measure of treatment benefit in PFS has stronger association with significant HR$_{OS}$ in favor of treatment -- HR$_{PFS}$ or (\%) median improvement in PFS? (b) Is there any other factor(s) (e.g., total number of deaths) that influence significance of HR$_{OS}$? (c) if yes, then how does this measure modify the association of treatment benefit in PFS with significant HR$_{OS}$?  The following variables were used as partitioning variables in the classification tree analysis: all three comparative PFS measures (HR$_{PFS}$, $\Delta$MED$_{PFS}$ and \%$\Delta$MED$_{PFS}$), total sample size  and total number of reported deaths. {\it Bagging} method \cite{breiman2001random} was applied to identify the most important partitioning variable(s). All statistical analyses were performed using R 3.0.2. A two-sided p value of $<$ .05 was considered statistically significant. ROC analysis was carried out using ``ROCR'' package \cite{ROCR}. Classification tree was constructed using ``rpart'' package \cite{rpart} and for {\it bagging} method we have used ``randomForest'' package \cite{randomforest}. 

\section*{Results}

\begin{table}
\begin{center}
\caption{ Summary of 74 comparisons (i.e. treatment to control comparisons) included in the meta-analyses } \label{Table2}
\begin{tabular}
{l|c}
\hline
Characteristics&\\
\hline
Study phase -- n(\%)&\\
\hspace{0.2in}Phase III&47 (63.5\%)\\
\hspace{0.2in}Phase II/IIB&19 (25.7\%)\\
\hspace{0.2in}Unknown&8 (10.8\%)\\
%&\\
%Randomization status -- n(\%)&\\
%\hspace{0.2in}Randomized&73 (98.6\%)\\
%\hspace{0.2in}Unknown&1 (1.4\%)\\
%&\\
Blinding status -- n(\%)&\\
\hspace{0.2in}Open&38 (51.4\%)\\ 
\hspace{0.2in}Blinded&15 (20.3\%)\\
\hspace{0.2in}Unknown&21 (28.3\%)\\
%&\\
Sample size (n=74)&\\
\hspace{0.2in}Median (Min, Max)&259 (41, 1349)\\
%&\\
Number of deaths (n=60)&\\
\hspace{0.2in}Median (Min, Max)&153 (19, 997)\\
%&\\
Type of control -- n(\%)&\\
\hspace{0.2in}Active&68 (91.9\%)\\ 
\hspace{0.2in}Placebo&3 (4.1\%)\\
\hspace{0.2in}Standard care&3 (4.1\%)\\
%&\\
Line of therapy -- n(\%)&\\
\hspace{0.2in}First line&49 (66.2\%)\\ 
\hspace{0.2in}2nd or beyond&25 (33.8\%)\\
%&\\
Increase in median PFS, $\Delta$MED$_{PFS}$ (n=72)&\\
\hspace{0.2in}Median (Min, Max)&1.60 (-0.50, 10.90)\\
%&\\
\% increase in median PFS, \%$\Delta$MED$_{PFS}$ (n=72)&\\
\hspace{0.2in}Median (Min, Max)&29.99 (-10.42, 294.60)\\
%&\\
HR in PFS, HR$_{PFS}$ (n=68)&\\
\hspace{0.2in}Median (Min, Max)&0.78 (0.24, 1.18)\\
%&\\
HR in OS, HR$_{OS}$ (n=63)&\\
\hspace{0.2in}Median (Min, Max)&0.85 (0.37, 1.49)\\
\hline
\end{tabular}
\end{center}
\end{table}

\begin{table}
\begin{center}
\caption{ Number of cases (i.e. treatment to control comparisons) reporting significant (at 5\% level) difference in PFS and OS time } \label{Table3}
\begin{tabular}
{l|c|c}
\hline
&\multicolumn{2}{c}{Overall survival (OS)}\\
\hline
Progression free survival (PFS)&HR$_{OS}$ significant&HR$_{OS}$ not significant\\
\hline
HR$_{PFS}$ significant&12&21\\
HR$_{PFS}$ not significant&4&37\\
\hline
\multicolumn{3}{l}{Level of significance is 5\%.}\\
\hline
\end{tabular}
\end{center}
\end{table}

\subsection*{Description}
We had a total of 74 treatment-to-control comparisons available from 65 publications  for the meta-analyses. The majority of these publications were published in the  Journal of Clinical Oncology (28; 43\%), Breast cancer research treatment (11; 17\%) and Annals of Oncology (9; 14\%), see Table \ref{Table1}. The majority (44, 68\%) of these studies recruited patients to treat as first line therapy. Forty-one (63\%) of these studies were phase III. In 60 studies comparison was made with active control, in 2 studies comparison was made with placebo and in remaining 3 studies standard care was used as comparator. \\

The characteristics of the  74 comparisons are summarized in Table \ref{Table2}. Of the 74 comparisons, 73 (98.6\%) were reported to be made in randomized set-up and 47 (63.5\%) were based on phase III trials.  Only 15 (20.3\%) comparisons were reportedly carried out in blinded fashion and blinding status was not reported for 21 (28.3\%) comparisons. The  median total sample size was 259 and the median number of deaths reported was 153. \textcolor{black}{The majority (91.9\%) of the comparisons included active control in the study, and in 66.2\% comparisons, treatment under investigation was the first line therapy.} The median HR$_{PFS}$ and HR$_{OS}$ were 0.78 and 0.85, respectively. Further, on average, median PFS time was increased by 1.60 months which translates to 29.99\% increase in median PFS time. \\

Of the 74 comparisons, significant (at 5\% level) HR$_{PFS}$ and HR$_{OS}$ were reported in 33 (44.6\%) and 16 (21.6\%) cases, respectively (see Table \ref{Table3}). The comparisons with significant HR$_{PFS}$ are 5.29 times more likely to have significant HR$_{OS}$ compared to the comparisons where HR$_{PFS}$ was not reported as significant. However, more importantly, only 36.4\% (12/33) of comparisons with significant HR$_{PFS}$ also reported significant HR$_{OS}$. 

\subsection*{Diagnostic accuracy (using ROC analysis)}

\begin{figure}
\centerline{%
\includegraphics [angle=0,width=100mm, height=180mm]{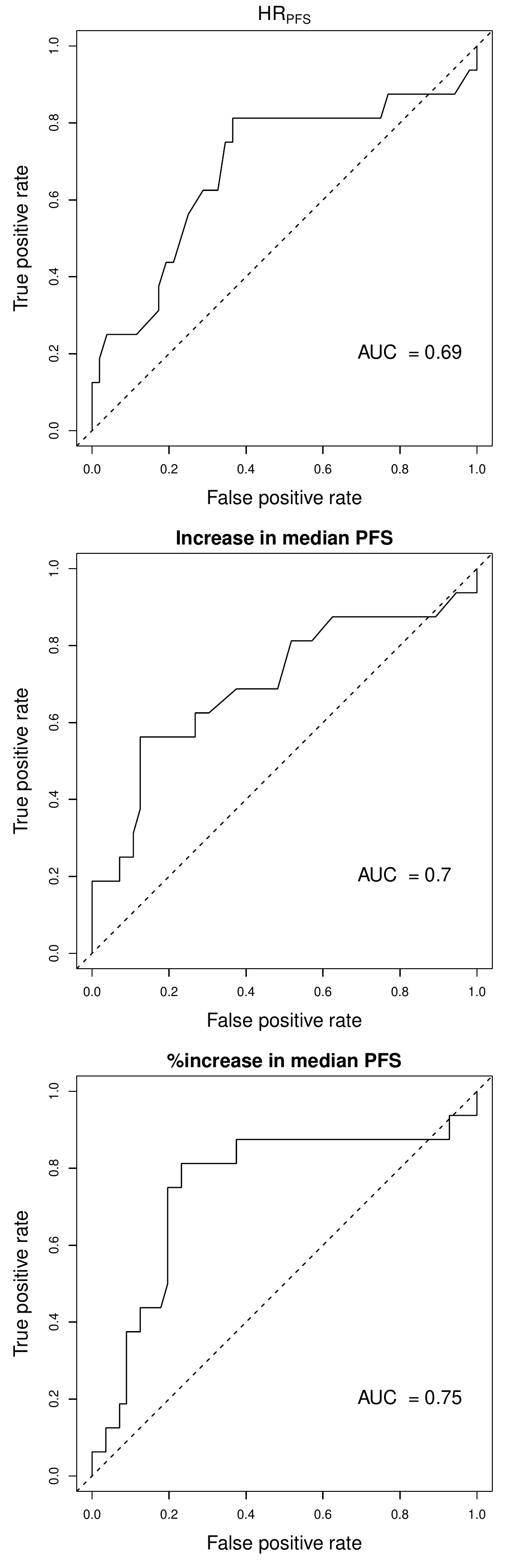}}
\caption{ROC curves using treatment effect on PFS as diagnostic measure for prediction of significant HR$_{OS}$ at 5\% level of significance. Treatment effect on PFS was assessed  using (a) hazard ratio (HR$_{PFS}$), (b) increase in median PFS ($\Delta$MED$_{PFS}$) and (c) \% increase in median PFS (\%$\Delta$MED$_{PFS}$). 
True positive rate was defined as proportion of correct prediction among the comparisons reporting significant HR$_{OS}$). False positive rate was defined as (\% of wrong prediction among the comparisons reporting non- significant HR$_{OS}$).}
\label{ROCfig}
\end{figure}

%Diagnostic accuracy of PFS measures (HR$_{PFS}$, $\Delta$MED$_{PFS}$ and \%$\Delta$MED$_{PFS}$) in predicting significant treatment effect on OS was assessed using receiver operating characteristics (ROC) curves and classification trees approach.  \\
ROC curves for each of HR$_{PFS}$,  $\Delta$MED$_{PFS}$ and \%$\Delta$MED$_{PFS}$ evaluating diagnostic accuracy to predict significant HR$_{OS}$ are displayed in Figure 1. AUC from ROC curves based on diagnostic measure of HR$_{PFS}$ (AUC=0.69) and $\Delta$MED$_{PFS}$ (AUC=0.70) were numerically close. However, \%$\Delta$MED$_{PFS}$  offers relatively better diagnostic accuracy with AUC as 0.75. From the ROC curve of \%$\Delta$MED$_{PFS}$ in Figure 1, the optimal cut-off point (according to Youden's index) was 44.83\%, for which the sensitivity (i.e. true positive rate) was 81.3\% and specificity (i.e. 1-false positive rate) 76.8\%. It can be  interpreted as follows: if we set a predictive rule  to classify the cases with improvement in median PFS greater than 44.83\% as producing significant HR$_{OS}$ subsequently, then 81.3\% of cases reporting significant HR$_{OS}$ will be correctly predicted and 76.8\% of cases reporting non-significant HR$_{OS}$ will be correctly predicted. Another cut-off point of interest could be 33.33\% for which the sensitivity and specificity were 87.5\% and  62.5\%, respectively.\\

\subsection*{Diagnostic accuracy (using classification tree)}

\begin{figure}
\centerline{%
\includegraphics [angle=0,width=150mm, height=70mm]{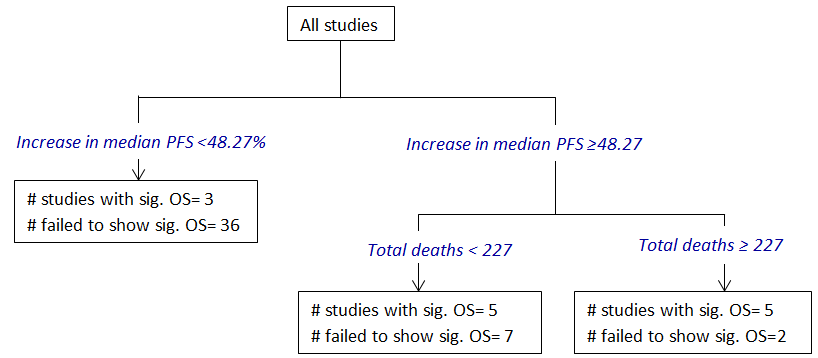}}
\caption{Classification tree for predicting significant treatment effect on OS}
\label{CART}
\end{figure}

\begin{figure}
\centerline{%
\includegraphics [angle=0,width=150mm, height=110mm]{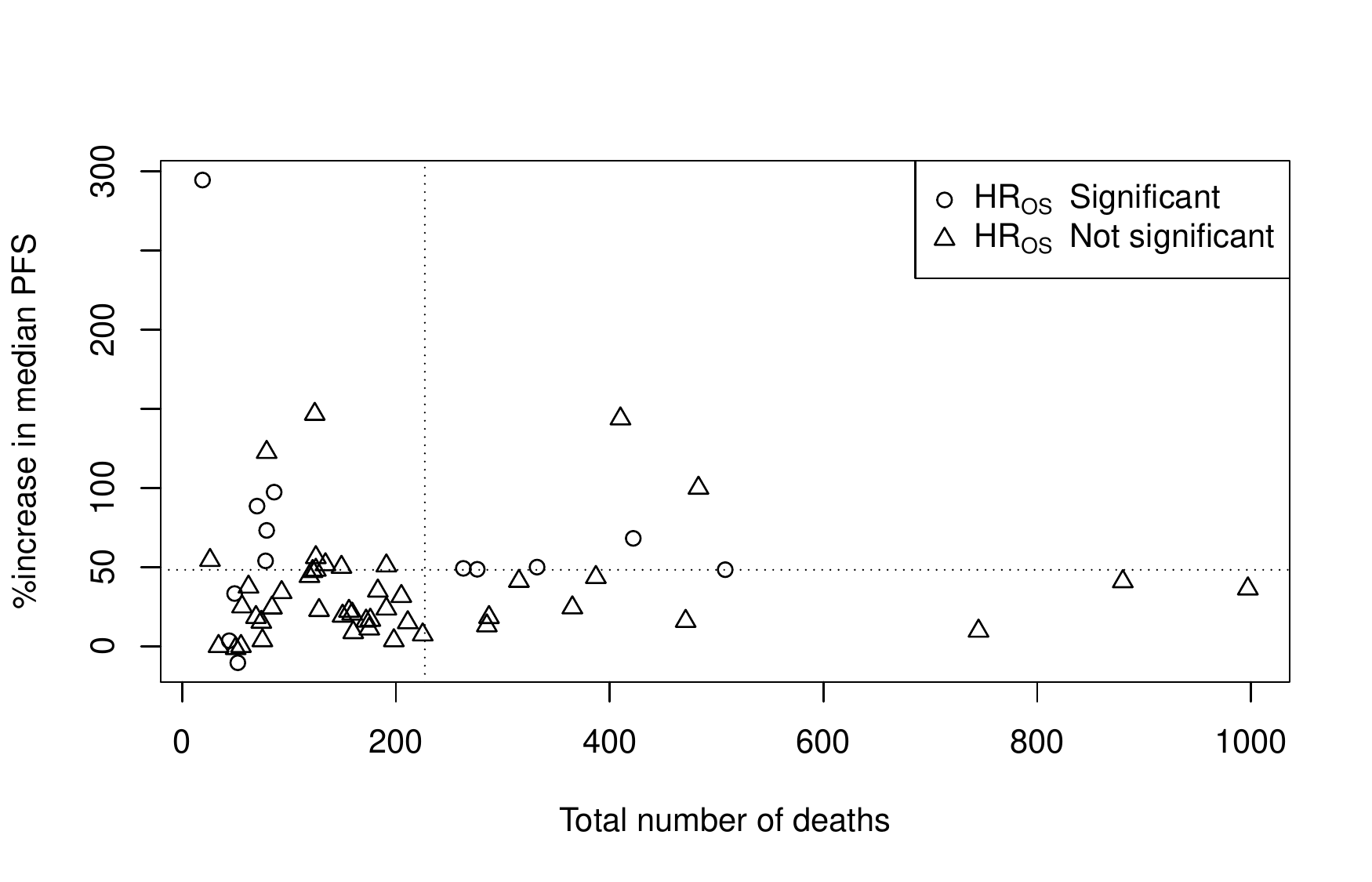}}
\caption{Display of comparisons with statistical significance status of HR$_{OS}$ in terms of \% increase in median PFS and total number of deaths}
\label{CART2}
\end{figure}

In classification tree approach, in addition to comparative PFS measures, number of deaths and sample size were also considered as predictor variables. Classification tree results based on 74 cases (i.e. comparisons) identified \%$\Delta$MED$_{PFS}$ and total number of deaths as diagnostic measures for significant HR$_{OS}$. Importantly, we found \%$\Delta$MED$_{PFS}$ as a more important predictor variable compared to HR$_{PFS}$ or $\Delta$MED$_{PFS}$. This is very much consistent with our findings observed in the analysis based on the ROC curve. The {\it bagging} method also suggested  \%$\Delta$MED$_{PFS}$ as the most important and total number of deaths as the second most important diagnostic measure for significant HR$_{OS}$.\\

Next, we performed classification tree analysis only on the 58 cases where information were available for both \%$\Delta$MED$_{PFS}$ and total number of deaths. The classification tree results are displayed in Figure~\ref{CART}. In 39 cases, increases in median PFS were shorter than 48.27\%; and only 3 of these cases showed significant HR$_{OS}$. \textcolor{black}{There were 19 cases with increases in median PFS reported at least 48.27\%. Of these 19 cases, in 12 cases total deaths reported were less than 227 and 5 of them reported significant HR$_{OS}$. In remaining 7 cases total deaths were 227 or more and 6 of them reported significant HR$_{OS}$. In Figure~\ref{CART2}, comparisons are displayed in terms of \%$\Delta$MED$_{PFS}$ and total number of deaths.  Figure~\ref{CART2} suggests that the treatment to control comparisons reporting statistically significant HR$_{OS}$ tend to have about 50\% or higher median PFS increase and total number deaths 227 or more. The findings of Figure~\ref{CART} and Figure~\ref{CART2} can be summarized as follows: There is only small chance that study would produce significant HR$_{OS}$ if increase in median PFS is less than 48.27\%. Trials with median PFS increase of at least 48.27\% seem to have better chance of producing statistically significant HR$_{OS}$, and having a total of 227 or more OS events further improves the likelihood of obtaining significant HR$_{OS}$.}

\section*{Discussion}
That a substantial improvement in PFS \textcolor{black}{may} be predictive of a corresponding difference in OS makes common sense. However, what is often not obvious is the magnitude of PFS difference that is required to be reasonably confident of observing a statistically significant HR$_{OS}$. This is crucial in late phase trials where therapeutic agents are being tested and the sponsor needs to decide whether the observed PFS difference could be predictive of a significant and clinically meaningful difference in OS and would merit a marketing authorization application (MAA). The ROC and classification tree analyses employed here are very well suited for such a determination. For example, the ROC approach gives us an overall assessment of diagnostic accuracy based on the AUC metric. On the other hand, the classification tree approach is helpful in identifying non-linear association and influence of other factor, such as total number of deaths, in an interpretable and visible manner. Both the approaches help us to choose an optimal operating point to guide the decision-making process.\\

%Venook and Tabernero \cite{venook2014progression} identified following reasons why  surrogacy of PFS for OS may be complicated in today's era: first, for immune-based therapies, tumor may in fact enlarge before shrinking. Second, patients are typically enrolled on studies with much lower volume of metastatic disease, and therefore, most patients are expected to live longer post-progression. Third, due to advances in treatment, patients have option to second or third line therapies. However due to these cross-over therapies, the concept of OS under any given treatment becomes less transparent. Therefore, prediction of OS based on PFS is very challenging in today's world compared to past decades. 
%
%Nevertheless, PFS continues to be the logical surrogate endpoint of OS, even though it's surrogacy is surrounded with uncertainty in some extent. The advantages of using PFS are that it can be determined much before OS and, unlike OS, it is not affected by subsequent therapies \cite{FDA2007cancer}. Therefore, PFS has historically been considered as surrogate endpoint from the standpoint of drug approvals \cite{johnson2003end, chmp2007methodological}.\\

Our study findings can be summarized as follows: First of all, $\Delta$MED$_{PFS}$ (i.e. percentage difference in median PFS) is a relatively better diagnostic predictor compared to HR$_{PFS}$. This is suggested by both ROC analysis and classification tree analyses. Secondly, higher \%$\Delta$MED$_{PFS}$ tends to be associated with significant HR$_{OS}$. However, our classification tree result suggests that higher number of deaths is also important in achieving significant HR$_{OS}$. The fact that the number of deaths influences the result of statistical testing of HR$_{OS}$ is very logical as deaths are considered as events in OS analysis and increased number of events improves the chance of statistical significance [i.e., power] in survival analyses.  
\textcolor{black}{Based on our study results,  the surrogate threshold effect (STE) in terms of \%$\Delta$MED$_{PFS}$ appears to be close to 50\% increase in median PFS (ROC analysis:  STE=44.83\%; Classification tree analysis: STE=48.27\%). This suggests the studies with 50\% improvement in median PFS are more likely to produce significant HR$_{OS}$ if the total number of OS events at the time analysis is 227 or more. This result can be useful in the context of breast cancer trials in at least two ways: first, a trial showing about 50\% increase in median PFS may serve as a useful indicator of statistically significant  HR$_{OS}$ while awaiting for OS data to mature. Secondly, if a prospective clinical trial plans to show statistically significant HR$_{OS}$, then that trial should plan for approximately 227 OS events. }\\

\textcolor{black}{An alternate approach to assessing diagnostic accuracy is the use of Positive predictive value (PPV) and Negative predictive value (NPV) curves introduced by Moskowitz and Pepe \cite{moskowitz2004quantifying}. While Positive predictive value (PPV) and negative predictive value (NPV) are great tools, these are not intrinsic measures of diagnostic accuracy, being dependent on the prevalence. In our case, pre-valence refers to prevalence of comparisons with statistically signficant HR$_{OS}$. In fact, Moskowitz and Pepe (2004), when commenting on the applicability of their method in the discussion section of their paper note "Study designs that result in a sample prevance that does not accurately reflect the true population prevalence will produce biased estimates of PPV and NPV". In their paper, they assume that the data arose from a cohort type design where the sample estimates would be valid. Since we are selecting a sample of studies from the whole population of studies using certain search criteria, there is no reason to assume that the sample prevalence would match the true population prevalence in this case. Also, given the relatively small number of positive studies, we believe that there is not enough data to draw a reliable PPV curve along the lines of Moskowitz and Pepe (2004). While we do acknowledge the intuitive appeal of the PPV and NPV curves, we would like to emphasize that the AUC would be a more appropriate metric for assessing diagnostic accuracy for our data.}\\
  
There have been numerous studies examining the strength of surrogacy and the ability of a surrogate measure such as PFS to predict OS. Most of the previous meta-analytic studies on metastatic breast cancer attempted to measure the association between improvement in  OS with improvement in PFS and clinical benefit rate (CBR) \cite{bruzzi2005objective, hackshaw2005surrogate,  miksad2008progression, sherrill2008relationship, burzykowski2008evaluation, ng2008correlation}. For example, Miksad {\it et al.} \cite{miksad2008progression} found only moderate correlation between HR$_{PFS}$ and HR$_{OS}$ ($R^2$ ranging between 0.35 to 0.59) in taxane and anthracycline based therapies in breast cancer patients. Recently, Amiri-Kordestani {\it et al.} (2016) \cite{amiri2016association} reported a moderate association between odds ratio (OR) of CBR and HR$_{PFS}$ ($R^2=0.52$), but failed to show any association with OR of CBR with HR$_{OS}$ ($R^2=0.01$) from 13 prospective studies submitted to FDA. %\textcolor{black}{Hackshaw {\it et al.} \cite{hackshaw2005surrogate} found a correlation of $0.87$ between treatment effects of PFS and HR$_{OS}$. Burzykowski {\it et al.} \cite{burzykowski2008evaluation} evaluated trial level surrogacy by fitting  simple (log-) linear regression analysis to model HR$_{OS}$ with ratio of median PFS time and reported $R^2$ of $0.48$.} \\
Our work takes a markedly different path. First, we are interested in assessing the association of treatment effect on PFS with the ultimate interest being the observation or non-observation of a significant HR$_{OS}$, whereas previous meta-analyses were mostly focused on exploring the association between the treatment effects on PFS and OS. Secondly, we have also considered PFS measures beyond HR$_{PFS}$ such as $\Delta$MED$_{PFS}$ (i.e. difference in median PFS) or \%$\Delta$MED$_{PFS}$ (i.e., percentage difference in median PFS). Thirdly, we have used fully data dependent non-parametric approaches like empirical ROC curves and classification tree  for the meta-analyses which, as far as we know, have not been used before in applications of this type. Fourthly, we have also included other factors such as total sample size and total number of deaths in evaluating the association between PFS measures and significance of HR$_{OS}$. Last but not the least, we have included a comprehensive list of all published studies since 2000 and hence the number of studies included in this current investigation is relatively higher than previous published meta-analyses.\\

Our study has its limitations. Not all breast cancer studies published during the period under consideration could be included, although the number is still quite high compared to previous meta analyses. The selection of studies was driven by common-sense, objective search criteria which were rigorously applied. Thus, only 74 treatment to control comparisons from 65 prospective published trials on breast cancer met the eligibility. The included trials were diverse in nature in terms of patient population considered -- some examples of patient population are patients with at least two prior chemotherapies, anthracycline or taxane resistant patients, post-menopausal patients,  HER2 positive patients, HER2 negative patients, just to name a few. \textcolor{black}{Further, the included studies were also diverse with regards to tumor type,  tumor stage and drug under investigation. We have reported only overall results from non-parametric analyses including all 74 studies; we did not have enough studies to assess surrogacy in each of the categories formed by the combination of these different factors.} Thus, the reported overall results may not be applicable in all set-up of breast cancer studies. There is also potential for publication bias. To minimize the publication bias, we have considered all the studies which were published and listed in PubMed database and met our search criteria. Lastly, we have not looked into other cancer indications, although we cannot think of any reason why the method would not be a useful tool in assessing surrogacy of PFS (or any other continuous time-to-event endpoint, for that matter) in other indications and settings.\\

To conclude, empirical ROC curves and classification trees could be useful tools for assessing how well treatment effect on PFS predicts weather HR$_{OS}$ would be statistically significant or not in breast cancer studies. 

%---------------------------------------------------------------------------------------------------------------------
%                          REFERENCES
%---------------------------------------------------------------------------------------------------------------------

%\bibliographystyle{unsrt}  %abbrv, acm, alpha, apalike, ieeetr, plain, siam and unsrt
%\bibliography{PFSvsOSbib}

%http://sites.stat.psu.edu/~surajit/present/bib.htm

\end{document}